\documentclass[%
 reprint,
 amsmath,amssymb,
 aps,
]{revtex4-2}

\usepackage{yfonts}
\usepackage{xcolor}
\usepackage{xcite}
\usepackage{braket}
\usepackage{graphicx}% Include figure files
\usepackage{dcolumn}% Align table columns on decimal point
\usepackage{bm}% bold math
\begin{document}

\preprint{APS/123-QED}

\title{
Unitary evolution and cosmic acceleration in Loop Quantum Cosmology}

\author{Omar Gallegos}
 \email{ogallegos@matmor.unam.mx}%Lines break automatically or can be forced with \\
 \affiliation{%
Centro de Ciencias Matemáticas, Universidad Nacional Autónoma de México, UNAM-Campus Morelia, A. Postal 61-3, Morelia, Michoacán C.P. 58090, México.\\
}%
\author{Tonatiuh Matos}%
 \email{tonatiuh.matos@cinvestav.mx}
\affiliation{%
Departamento de Física, Centro de Investigación y de Estudios Avanzados del IPN, A.P. 14-740, 07000 CDMX, México.\\
}%

\author{Hugo A. Morales-Técotl}
 \email{hugo@xanum.uam.mx}
\affiliation{
 Departamento de Física, Universidad Autónoma Metropolitana Iztapalapa, Av. Ferrocarril San Rafael Atlixco, Núm. 186, Col. Leyes de Reforma 1 A Sección, Alcaldía Iztapalapa, C.P. 09310, Ciudad de México, México.\\
}%

\date{\today}% It is always \today, today,

\begin{abstract}

Loop quantum cosmology was shown to interpolate between de Sitter and FLRW Universe phases through a bounce by including  Euclidean and Lorentzian terms of the Hamiltonian constraint with weight one -that corresponding to classical General Relativity. Unitary evolution required self-adjoint extensions of the constraint and a Planckian cosmological constant was obtained. Independent work took a positive weight to get a cosmological constant with the observed value, without considering unitarity. 
In this work we address the unitary evolution of the model for arbitrary weight. For non positive weight parameter unitary holds but for positive values self-adjoint extensions are required. To encompass observations the extensions here provided are mandatory. These are implemented in a propagator. Finally, we discuss our results and perspectives.

\end{abstract}

\maketitle

%\tableofcontents

\color{black} 

\section{Introduction}
\label{section:introduction}

The observational evidence  \cite{Planck:2018nkj,Planck:2018vyg} in favor of a, possibly depending on time \cite{DESI:2024mwx}, accelerated expansion of our universe takes us back to the question as to what is its origin \cite{Weinberg:1988cp}. 
In its simplest form this amounts to the so-called cosmological constant problem. However, as it is well known, a simple minded combination of field theory and gravity leads to a value for this parameter 120 orders of magnitude off the current observations. Given this state of affairs the possibility has been considered to find a solution for this challenge within quantum gravity. Amongst different proposals, Loop Quantum Gravity (LQG) stands forward in regard to the inclusion of a background independent discrete structure \cite{Rovelli:2004tv,Thiemann:2007pyv}, that has succeeded in the study of black holes \cite{Morales-Tecotl:2018ugi} and cosmological models \cite{Bojowald:2001xe,Bojowald:2005epg,Ashtekar:2003hd,Ashtekar:2006es,Ashtekar:2006wn,Pawlowski:2011zf}, in, for instance, replacing classical singularities by quantum bounces. 

Using LQG for homogeneous models is dubbed Loop Quantum Cosmology (LQC) \cite{Bojowald:2001xe}. In this case
the two terms entering the Hamiltonian constraint: Euclidean and Lorentzian, turn out to be proportional classically, and quantization originally focused on the former \cite{Bojowald:2001xe,Bojowald:2005epg,Ashtekar:2003hd,Ashtekar:2006es,Ashtekar:2006wn,Pawlowski:2011zf,Kaminski:2007gm,Kaminski:2009pc,Bentivegna:2008bg, Martin-Benito:2009htq, Ashtekar:2007em, Szulc:2006ep,Szulc:2007uk,Kaminski:2007gm}. Nevertheless, 
the possibility of quantizing the Lorentzian term along the lines of LQG \cite{Thiemann:2007pyv} was pursued
\cite{Yang:2009fp,Zhang:2021zfp,Assanioussi:2018hee,Assanioussi:2019iye,Yang:2022aec} yielding interpolation between a de Sitter and FLRW phases; effectively a cosmological constant. Either one or the other of the following  unappealing features hold for these models: (i) the scale of the cosmological constant is Planckian, or, (ii) no unitary evolution is ensured since its generator is not self-adjoint.

The cosmological constant issue has been examined within LQC, by adopting a Friedman-Lemaître-Robertson-Walker (FLRW) model, which describes a homogeneous and isotropic Universe with a massless scalar field $\phi$ as a clock. At first an explicit cosmological constant was considered in the Hamiltonian constraint, with only the Euclidean term \cite{Ashtekar:2006wn, Pawlowski:2011zf,Kaminski:2007gm,Kaminski:2007gm,Kaminski:2009pc, Bentivegna:2008bg, Martin-Benito:2009htq}. Here the Hamiltonian operator is not self-adjoint, but admits self-adjoint extensions \cite{Pawlowski:2011zf,Kaminski:2009pc}. On the other hand, some works \cite{Zhang:2021zfp,Assanioussi:2018hee,Assanioussi:2019iye,Yang:2022aec} concentrated on a flat FLRW cosmological model including both terms in the Hamiltonian constraint without cosmological constant. In this case the inclusion of the Lorentzian term yields an effective Planckian cosmological constant \cite{Assanioussi:2018hee,Assanioussi:2019iye}. To get an effective cosmological constant with the observable value, it was proposed to include a parameter $\lambda$ {\it weighting} differently the Euclidean and Lorentzian terms of the Hamiltonian constraint \cite{Zhang:2021zfp}.

The de Sitter branch emerging from the bounce is described asymptotically by the acceleration equation
\begin{equation}
\frac{\ddot{a}}{a}=
     - \frac{4 \pi GC_{\gamma}}{3}(\rho+3 P)+\frac{\Lambda_{\mathrm{eff}}}{3},
\end{equation}
with $C_{\gamma}=\left(\frac{1-5 \lambda \gamma^2}{1+\lambda \gamma^2}\right)$ and $\Lambda_{\mathrm{eff}} = \frac{3\lambda }{\left(1+ \lambda\gamma^2\right)^2 \Delta}$, the effective cosmological constant.
$a$ is the scale factor, dot derivative w.r.t. $t$, the cosmic time,   $\gamma=0.2375$ is the Barbero-Immirzi parameter \cite{Meissner:2004ju,Ghosh:2004rq,Rovelli:1996dv,Ashtekar:1997yu} and
$\rho$ and $P$ are energy density and pressure of matter. $\Delta$ is the gap of area in LQG of order $\ell_P^2$, with $\ell_P$ Planck's length. In the regime dominated by $\Lambda_{\mathrm{eff}}$ the universe presents positive acceleration whenever $\lambda>0$. Unfortunately, the unitary evolution or self-adjointness of the Hamiltonian constraint was not taken into account.

In this work, to ensure unitary evolution, we study the self-adjointness of the flat FLRW Hamiltonian constraint with a weight parameter between the Euclidean and the Lorentzian terms. We use the soluble form of this model and apply the deficiency indices method \cite{Reed:1972uy,Reed:1975uy,Simon98theclassical,Kato:1995,gitman:2012}, in order to combine both effects of the weight parameter and the self-adjoint character of the Hamiltonian.

The necessary boundary condition to archive self-adjointness can be implemented into a propagator and not only in the spectrum of the gravitational operator as in previous work. This may contribute to the construction of an effective dynamics leading to physical considerations.

This work is organized as follows. Section \ref{section:Classical model} presents the classical cosmological model (homogeneous, isotropic and flat) with Hamiltonian constraint in terms of the Ashtekar-Barbero variables including a weight parameter between the Euclidean and the Lorentzian terms. In Section \ref{section:LQC} the quantum model based on Thiemann's regularization is given. A previous representation is adopted to facilitate the self-adjointness analysis. In Section \ref{section:Self-adjointness}, the deficiency indices method is applied to find the possible self-adjoint extensions to define the unitary evolution. In Section \ref{section:Eff model} the implementation of self-adjoint extensions into the cosmic propagator is provided showing the effect of the boundary condition to be fulfilled. Finally, in Section \ref{section:Conclusion}, we discuss our results as well as possible future work.

\section{Classical model}
\label{section:Classical model}
To describe the cosmological model we consider a spacetime manifold $M=\mathbb{R}\times \Sigma$, where $\Sigma$ is 3D  \cite{Rovelli:2004tv,Thiemann:2007pyv}. The dynamics in the Hamiltonian formulation gets expressed in terms of the Gaussian, diffeomorphisms and Hamiltonian constraints. However, for homogeneous and isotropic FLRW the Hamiltonian constraint is the only one not trivial, and that is what we will focus on. Furthermore, the Ashtekar-Barbero variables $(A^i_a(x), E^a_j(x))$ coordinatize the classical phase space. Here, $A^i_a(x)$ is a $SU(2)$ connection and $E^a_j(x)$ is the densitized triad, which are defined as 
\begin{align}
    \label{ashtekar connection}
    A^i_a=\Gamma^i_a+\gamma K^i_a, \\
    \label{densitized triad}
    E^a_i=\sqrt{q}e^a_i,
\end{align}
where $e^a_i$ is the triad, $q^{ab}=e^a_i e^b_j\delta^{ij}$ and $\Gamma_a^i$ is the spin connection compatible with the triad, while $K^i_a$ is the extrinsic curvature and $q$ denotes the determinant of the spatial metric $q_{ab}$ of $\Sigma$. In addition, $F_{ab}^i$ is the curvature of $A^i_a$, namely,
\begin{equation}
    \label{curvature connection}
    F_{ab}^i=2\partial_{[a}A^i_{b]}+\epsilon^i_{jk}A_a^jA_b^k.
\end{equation}
Using these ingredients the Hamiltonian constraint becomes  \cite{Zhang:2021zfp}
\begin{eqnarray}
     \label{geo Hamiltonian constraint}
    H_g(N)&=&\lambda H^E(N)-2(1+\lambda\gamma^2)H^L(N)\nonumber\\
    &+&\dfrac{-1+\lambda}{16\pi G}\int_\Sigma \mathrm{d}^3x N\sqrt{q} ^{(3)}R,\\
        \label{Euclidean term} 
    H^E(N)&=&\dfrac{1}{16\pi G}\int_\Sigma \mathrm{d}^3x NF^j_{ab}\dfrac{\epsilon_{jkl}E^a_kE^b_l}{\sqrt{q}},\\
    \label{Lorentzian term}
    H^L(N)&=&\dfrac{1}{16\pi G}\int_\Sigma \mathrm{d}^3x N\epsilon^j_{mn}K_a^m K_b^n\dfrac{\epsilon_{jkl}E^a_kE^b_l}{\sqrt{q}},
\end{eqnarray}
where $G$ is the Newtonian constant, $N$ is the lapse function and $^{(3)}R$ is the 3-dimensional curvature of $\Sigma$. Furthermore, $\lambda$ is introduced as a weight parameter between the Euclidean $H^E(N)$ and the Lorentzian $H^L(N)$ terms. In what follows we consider the spatially flat case, where $^{(3)}R$ is zero. 

Because of the spatial homogeneity of the flat FLRW metric 
\begin{eqnarray}
    \mathrm{d}s^2=-N^2\mathrm{d}t^2+a^2\mathrm{d}\textbf{x}^2,
\end{eqnarray}
we need to introduce an elementary cell of volume $\mathcal{V}$ in the spatial manifold $\mathbb{R}^3$ in order to make meaningful the symplectic structure and some observables. The symmetric canonical variables are \cite{Bojowald:2001xe,Ashtekar:2003hd,Ashtekar:2006wn,Ashtekar:2007em}
\begin{eqnarray}
    \label{cosmo Ashtekar variables}
    A^i_a=c\ ^0\omega_a^iV_0^{-1/3}, \ \ E^b_j=pV_0^{-2/3}\sqrt{^0q} \ ^0e^b_j,
\end{eqnarray}
where $^0q_{ab}$ is a fiducial metric on $\mathbb{R}^3$ and its determinant is written as $^0q$. Furthermore, the orthonormal triad and co-triad ($^0e^a_i,^0\omega^i_a$) on $\mathcal{V}$ are such that $^0q_{ab}=\ ^0\omega_a^i \ ^0\omega_b^i$. In addition, $V_0$ denotes the volume of $\mathcal{V}$ measured by $^0q_{ab}$. $c$, $p$ are the canonical variables, which are only functions of cosmological time $t$. To include a physical clock we use a homogeneous scalar field $\phi$. This leads to the addition of a matter term to the Hamiltonian constraint
\begin{eqnarray}
    \label{matter Hamiltonian}
    H_\phi&=&\dfrac{p_\phi^2}{2V},
\end{eqnarray}
where $V=|p|^{3/2}$ is the physical volume and $p_\phi$ is the momentum of the massless scalar field $\phi$, which is a monotonic function of the cosmic time $t$. Hence, the phase space is expressed in terms of ($c,p$) and ($\phi, p_\phi$), which satisfy the following Poisson brackets \begin{eqnarray}
    \label{c p poisson bracket}
    \{c,p\}=\dfrac{8\pi G\gamma}{3}, \ \ \ \{\phi,p_\phi\}=1.
\end{eqnarray}

In order to avoid large quantum gravity effects in semiclassical regimes it is convenient to adopt improved variables ($b,v$) given by \cite{Ashtekar:2006wn}
\begin{eqnarray}
    \label{b v variables realation}
    v=2\sqrt{3} \textrm{sgn}(p)\bar{\mu}^{-3}, \ \ \ b=\bar{\mu}c,
\end{eqnarray}
where $\bar{\mu}=\sqrt{\Delta/|p|}$. Being  $\Delta=4\sqrt{3}\pi\gamma G\hbar$ the gap of area, that is the minimum non-zero eigenvalue of the area operator and $\hbar$ is Planck's constant. They obey
\begin{eqnarray}
    \label{bv poisson brackets}
    \{b,v\}=\dfrac{2}{\hbar}.
\end{eqnarray}
These are related to the scale factor in the following fashion \cite{Pawlowski:2011zf}
\begin{eqnarray}
    \operatorname{sgn} (v) v= \frac{a^3 V_0}{2 \pi \gamma \sqrt{\Delta} G\hbar},\quad b= \gamma \sqrt{\Delta} \frac{1}{a} \frac{\mathrm{d} a}{\mathrm{d} t}.
\end{eqnarray}

The Euclidean (\ref{Euclidean term}) and the Lorentzian (\ref{Lorentzian term}) terms are classically proportional to each other. For spatially flat FLRW this is expressed as
\begin{eqnarray}
 H^{E}(1)=\dfrac{2}{\gamma^2} H^{L}(1)=\frac{3 \gamma \hbar}{4 \sqrt{\Delta}} b^2|v|.
\end{eqnarray}
To include at the quantum level the Lorentizian term amounts to quantize separately the two term in (\ref{Euclidean term}) and (\ref{Lorentzian term}) taking as a guide full LQG \cite{Thiemann:2007pyv}. We proceed to do this next.

\section{Loop Quantum Cosmology}
\label{section:LQC}
The kinematical Hilbert space for the geometry part is given by $\mathcal{H}_{kin}^{g}=L^2(\mathbb{R}_{Bohr},\textrm{d}\mu_H)$ where $\mathbb{R}_{Bohr}$ is the compactification of the real line and $\mathrm{d}\mu_H$ is the Haar measure on it \cite{Ashtekar:2003hd}. On the other hand, for the scalar field we have $\mathcal{H}^\phi_{kin}=L^2(\mathbb{R},\mathrm{d}\mu)$ where $\mathbb{R}$ is the real line and $\mathrm{d}\mu$ is the Lebesgue measure. Thus, the full Hilbert space is $\mathcal{H}_{kin}=\mathcal{H}^g_{kin}\otimes\mathcal{H}^\phi_{kin}$. Taking the (discrete) $v-$representation of the basic operators $\widehat{\mathcal{N}}=\widehat{e^{i b/2}}$ and $\widehat{v}$ given by
\begin{align}
    \widehat{\mathcal{N}}\ket{v}=\ket{v+1}, \quad  \widehat{v}\ket{v}=v\ket{v}, \quad \braket{v'|v}=\delta_{v',v}
\end{align}
where $v,v'\in\mathbb{R}$. In the case of the matter field  we have 
\begin{align}
    \widehat{\phi}\ket{\phi}=\phi\ket{\phi}, \quad  \widehat{p}_\phi\ket{\phi}=-i\partial_\phi\ket{\phi}, \quad \braket{\phi'|\phi}=\delta(\phi'-\phi).
\end{align}
A basis of $\mathcal{H}_{kin}$ is $\ket{v,\phi}=\ket{v}\otimes\ket{\phi}$ and the inner product here is \cite{Pawlowski:2011zf}
\begin{equation}
\label{inner product}
    \left\langle\psi\mid\psi^{\prime}\right\rangle=\int_{\mathbb{R}}\mathrm{d}\phi\sum_{v \in \mathbb{R}}  \bar{\psi}(v,\phi) \psi^{\prime}(v,\phi).
\end{equation}

As for the dynamics we need to quantize the complete constraint $H_g+H_\phi$ (eqs. (\ref{geo Hamiltonian constraint}) and (\ref{matter Hamiltonian})). We apply Thiemann's regularization \cite{Thiemann:2007pyv,Yang:2009fp,Zhang:2021zfp,Assanioussi:2018hee,Assanioussi:2019iye} together with a convenient approximation that makes the model soluble \cite{Ashtekar:2007em,Assanioussi:2018hee,Assanioussi:2019iye}. From here on we do not make explicit the $\phi$ dependence unless it is strictly required. The result can be expressed as
\begin{eqnarray}
 \label{LQC full theta operator weight}
&&\widehat{\Theta}_\lambda=\mathbb{I}\otimes\partial_\phi^2+\widehat{\Theta}_{\lambda,g}\otimes\mathbb{I}, \\ 
\label{flat hamiltonian operator}
&&\widehat{\Theta}_{\lambda,g}\psi(v)=\dfrac{3\pi G\gamma^2}{4}\sum_{j=-2}^2 M_{4j}\psi(v+4j),
\end{eqnarray}
where $M_{\pm 4}=-\lambda f_{\pm 4}(v)$, $M_{\pm 8}=\xi_\lambda f_{\pm 8}(v)$ and $M_0=-2(\xi_\lambda-\lambda)f_0(v)$ with $f_{4j}(v)=\sqrt{|v(v+4j)|}|v+2j|$ and
\begin{eqnarray}
    \xi_\lambda=\dfrac{1+\lambda\gamma^2}{4\gamma^2}.
\end{eqnarray}

Although originally unitary evolution in LQC was studied in the $v$-representation \cite{Pawlowski:2011zf,Kaminski:2009pc, Kaminski:2007gm,Assanioussi:2018hee,Assanioussi:2019iye}, the difficulty in dealing with difference operators and the corresponding norms of states has led to the more convenient $b-$representation with differential operators and squared integrable functions. We will do this in preparation for our self-adjointness analysis of the Hamiltonian constraint of next section. 

Let us consider the discrete Fourier transformation \cite{Ashtekar:2006wn,Ashtekar:2007em,Pawlowski:2011zf,Assanioussi:2018hee,Assanioussi:2019iye} 
\begin{equation}
    \label{Fourier trasformation}
\Tilde{\psi}(b)=\left[\mathcal{F}\psi\right](b)=\sum_{v\in\mathcal{L}_4}\dfrac{1}{\sqrt{|v|}}\psi(v)e^{ivb/2},
\end{equation}
where $\mathcal{L}_4$ is the superselected sector appeared in (\ref{flat hamiltonian operator}); notice a parity reflection symmetry holds $\Tilde{\psi}(b)=\Tilde{\psi}(\pi-b)$, namely, it has period $\pi$ and $b\in [0,\pi]$. Now, the basic operators act in the following way 
\begin{equation}
    \widehat{v}\ket{b}=2i\partial_b\ket{b}, \ \ \mathrm{and} \ \ \widehat{\mathcal{N}}\ket{b}=e^{-i b/2}\ket{b}.
\end{equation}
In this case, (\ref{flat hamiltonian operator}) takes the form
\begin{equation}
\label{b full hamiltonian operator}
  \widehat{\Theta}_{\lambda,g}\Tilde{\psi}(b)=12\pi G\gamma^2\left[\lambda(\sin(b)\partial_b)^2-\xi_\lambda(\sin(2b)\partial_b)^2\right]\Tilde{\psi}(b).  
\end{equation} 
By referring to (\ref{geo Hamiltonian constraint}) we can see $\widehat{\Theta}_{\lambda,g}$ reduces to the Euclidean term when $\lambda=-1/\gamma^2$ \cite{Ashtekar:2006wn,Kaminski:2007gm,Ashtekar:2007em}, it becomes the Lorentzian term when $\lambda=0$ and when $\lambda=1$ it takes the form dictated by classical GR \cite{Yang:2009fp,Assanioussi:2018hee,Assanioussi:2019iye}. 

Following \cite{Assanioussi:2018hee,Assanioussi:2019iye,Pawlowski:2011zf}, a further change of representation facilitates the application deficiency indices method to be applied in the next section. This amounts to give $\widehat{\Theta}_{\lambda,g}$ (\ref{b full hamiltonian operator}) a Klein-Gordon like form, and there are three possibilities depending on $\lambda\lessgtr 0 $.

\textit{Case $\lambda\leq 0$}.
Starting with $\lambda<0$ the $x-$representation is defined by \cite{Kowalczyk:2022ajp}
\begin{equation}
\label{x(b) lambda<0}
   x(b) =-\operatorname{arctanh}\left(\frac{\cos (b)}{\left(1+\lambda \gamma^2\right) \sin ^2(b)-1}\right).
\end{equation}
Let us notice $x(b)\in \mathbb{R}$ and $\lim _{b \rightarrow 0} x(b)=-\infty, \quad \lim _{b \rightarrow  \pi} x(b)=+\infty.$ The operator takes the form 
\begin{equation}
\label{KG op}
   \widehat{\Theta}_{\lambda\leq 0,g}\psi(x)=-12\pi G\partial_x^2\psi(x).
\end{equation}
For $\lambda=0$, the relation between $x$ and $b$ is
\begin{eqnarray}
\label{x(b) lambda=0}
   x(b)=\ln |\tan (b)|,
\end{eqnarray}
here $x\in \mathbb{R}$ and $\lim _{b \rightarrow 0} x(b)=-\infty,\quad \lim _{b \rightarrow \pi} x(b)=-\infty,\quad \lim _{b \rightarrow \pi/2^{\pm}}x(b)=+\infty.$ The operator takes the same form as in (\ref{KG op}) and the corresponding eigenvalue problem yields the following eigenfunctions and spectrum for $\lambda\leq 0$
\begin{eqnarray}
\label{psi lneg0}
    \psi_k (x) &=& \zeta\cos(kx),\\
    k &=& \sqrt{\dfrac{\sigma}{12\pi G}}, 
\end{eqnarray}
where $\zeta=4/\sqrt{|k|}$ and $\sigma$ is a positive eigenvalue  \cite{Pawlowski:2011zf,Assanioussi:2018hee,Assanioussi:2019iye}. In this case the spectrum is continuous.

\textit{Case $\lambda> 0$}.
We have the $x-$representation given by
\begin{eqnarray}
\label{x(b) lambda>0}
    x(b)=\left\{\begin{array}{l}
\frac{1}{2} \ln B_1-\frac{\pi}{2}, \quad b \in\left(0, b_0\right), \\
-\arctan B_2, \quad b \in\left(b_0, \pi-b_0\right), \\
\frac{1}{2} \ln B_1+\frac{\pi}{2}, \quad b \in\left(\pi-b_0, \pi\right) ,
\end{array}\right.
\end{eqnarray}
where
\begin{eqnarray}
    &&B_1(b)=1-\frac{2 \sqrt{1-\left(1+\lambda \gamma^2\right) \sin ^2(b)}}{\cos (b)+\sqrt{1-\left(1+\lambda \gamma^2\right) \sin ^2(b)}},\\
    &&B_2(b)=\frac{\cos (b)}{\sqrt{\left(1+\lambda \gamma^2\right) \sin ^2(b)-1}},
\end{eqnarray}
with 
\begin{equation}
    b_0=\arccos\left(\sqrt{\dfrac{\lambda\gamma^2}{1+\lambda\gamma^2}}\right).
\end{equation}
Here $x\in \mathbb{R}$, $\lim _{b \rightarrow 0} x(b)=-\infty,\quad \lim _{b \rightarrow \pi} x(b)=+\infty,\quad x(b=\pi/2)=0,\quad
x\left(b_0\right)=-\frac{\pi}{2},\quad x\left(\pi-b_0\right)=\frac{\pi}{2}$, and the operator becomes 
\begin{equation}
\label{sgn KG op}
   \widehat{\Theta}_{\lambda>0,g}\psi(x)=-12\pi G\text{sgn}(|x|-x_0)\partial_x^2\psi(x),
\end{equation}
where $x_0=-x(b_0)$.

At this point, we have two observations: First, the model presents two different regimes depending on whether $\lambda\leq 0$ or $\lambda>0$, interpolating between Euclidean ($\lambda=-1/\gamma^2$), Lorentzian ($\lambda=0$) and the one dictated by classical GR ($\lambda=1$). Second,  the $\text{sgn}(x)$ in (\ref{sgn KG op}) indicates a boundary condition is required such that at $x=\pm \pi/2$ wave functions are continuous but not differentiable. This property will be crucial in defining self-adjoint extensions as we are about to see. 

\section{Unitary evolution}
\label{section:Self-adjointness}
Self-adjointness in quantum mechanics is relevant to characterize physical observables. In particular, for a Hamiltonian it ensures unitary time evolution conserving probability and scalar product. Mathematically, a self-adjoint operator is defined as hermitian (symmetric) $\widehat{\Theta}_{\lambda,g}=\widehat{\Theta}_{\lambda,g}^\dagger$, with $\widehat{\Theta}_{\lambda,g}^\dagger$ the hermitian adjoint of $\widehat{\Theta}_{\lambda,g}$ and, moreover, with corresponding domains equal $D(\widehat{\Theta}_{\lambda,g})=D(\widehat{\Theta}_{\lambda,g}^\dagger)$. Thus, there exist some hermitian operators that are not self-adjoint. Nevertheless, it is possible for them to define extensions for their domains to make them self-adjoint (c.f. \cite{Reed:1972uy, Reed:1975uy,Simon98theclassical,Kato:1995,gitman:2012}).

The deficiency indices method due to von Neumann can be used to study the self-adjointness of the Hamiltonian operator $\widehat{\Theta}_{\lambda\leq 0,g}$ and $\widehat{\Theta}_{\lambda>0,g}$, eqs. (\ref{KG op}) and (\ref{sgn KG op}). As a first step we need to identify the deficiency subspaces $\mathcal{K}^{\pm}$, which are the spaces of the normalizable solutions $\psi^{\pm}$ of the following complex eigenvalue equations 
\begin{eqnarray}
\label{deciency equation}
\widehat{\Theta}_{\lambda,g}\psi^\pm(x)=\pm 24\pi G i\psi^\pm(x).
\end{eqnarray}

This method focuses on the dimension of these deficiency subspaces, $n_\pm=\textrm{dim}(\mathcal{K}^\pm)$, called deficiency indices. The possibilities can be divided into three cases: i) If $n_+=n_->0$, the operator admits a family of possible extensions. ii) If $n_+=n_-=0$, there exists a single extension and the operator obtained hereby is called essentially self-adjoint. iii) If $n_+ \neq n_-$, it admits no extensions.

\textit{Case $\lambda\leq 0$}. Applying this method to (\ref{KG op}) yields
\begin{eqnarray}
    \psi^{\pm}(x)=c_{1}\exp \left(\frac{1\pm i}{\sqrt{2}} x\right)+c_{2}\exp \left(-\frac{1\pm i}{\sqrt{2}} x\right),
\end{eqnarray}
where $c_{1}$, $c_{2} \in \mathbb{C}.$ These solutions are not normalizable in the Hilbert space $\mathcal{H}_{kin}^g$, and hence, $c_1=c_2=0$. The deficiency sub-spaces contain only the null vector, that is, $n_\pm=\mathrm{dim}(\mathcal{K^\pm})=0$. With this result, we can conclude that the Hamiltonian operator $\widehat{\Theta}_{\lambda\leq 0,g}$ is essentially self-adjoint. This contains the Euclidean case ($\lambda=-1/\gamma^2$) \cite{Kaminski:2007gm,Kaminski:2009pc}. 
and the evolution is unitary for eigenstates $ \psi_{k}(x,\phi)$ (\ref{psi lneg0}) 
\begin{equation}
    \psi_{k}(x,\phi)=e^{i(\phi-\phi_0)\sqrt{|\widehat{\Theta}_{\lambda\leq 0,g}|}}\psi_{k}(x,\phi_0),
\end{equation}
where the square root is defined for the positive part of the operator $\widehat{\Theta}_{\lambda\leq 0,g}$, eq. (\ref{KG op}).

\textit{Case $\lambda> 0$}. The analysis of the deficiency indices parallels that of  \cite{Assanioussi:2018hee,Assanioussi:2019iye}. The deficiency subspaces corresponding to (\ref{sgn KG op}) are given in terms of 
\begin{eqnarray}
\label{deficy indices function}
   \psi^{ \pm}(x)=\zeta \begin{cases}\left(e^\pi-1\right) e^{( \pm i-1)|x|}, & |x|>\pi / 2, \\ e^{(1 \pm i) x}+e^{-(1 \pm i) x}, & |x| \leq \pi / 2,\end{cases}
\end{eqnarray}
Since the solutions (\ref{deficy indices function}) are normalizable, the deficiency subspaces $\mathcal{K}^\pm$ are not empty and $n_+=n_-=1$. In this case the operator admits a family of self-adjoint extensions, which is associated with a unitary transformation $U^\alpha:\mathcal{K}^+\xrightarrow{} \mathcal{K}^{-}$ where each extension corresponds to an element of $U(1)$, which acts as
\begin{equation}
    \label{Unitary transf}
    U^\alpha:\psi^+\rightarrow e^{i\alpha}\psi^-.
\end{equation}
The uniparametric family of self-adjoint extensions is labeled by the parameter $\alpha\in(0,2\pi]$. Furthermore, from Theorem X.2 in \cite{Reed:1975uy} applied to the extended domain $D_\alpha$ of the Hamiltonian operator, and corresponding extension $\widehat{\Theta}_\alpha$, we have
\begin{eqnarray}
\label{extended domain}
D_\alpha=\{\psi+\psi^+ +U^\alpha\psi^+;\ \psi\in D,\ \psi^\pm\in\mathcal{K}^\pm \},\\
\widehat{\Theta}_\alpha(\psi+\psi^+ +U^\alpha\psi^+)=\hat{\Theta}_\alpha\psi+i\psi^+ -iU^\alpha\psi^+,
\end{eqnarray}
where $D= L^2(\mathbb{R}_{Bohr},\textrm{d}\mu_H)$. 

To characterize the self-adjoint extensions it is useful to define $\psi^\alpha=\psi^++U^\alpha\psi^+$ having the explicit form
\begin{eqnarray}
\label{psi extended}
    \psi^\alpha(x)=\zeta\left\{\begin{array}{rcl}
    a(x),    & |x|>\pi/2, \\
    b(x),   & |x|\leq \pi/2, 
    \end{array} \right.
\end{eqnarray}
where functions $a(x)$ and $b(x)$ are given by 
\begin{eqnarray}
    &&a(x)=4\cosh \left(\frac{\pi}{2}\right) e^{-|x|+\frac{\pi}{2}+i\frac{\alpha}{2}} \cos \left(|x|-\frac{\alpha}{2}\right),\\
    &&b(x)=2e^{i\alpha / 2} \left[e^x \cos \left(x-\frac{\alpha}{2}\right)+ e^{-x}\cos \left(x+\frac{\alpha}{2}\right)\right].
\end{eqnarray}
The boundary condition consisting of the discontinuity for the wave function at $x=\pm \pi/2$ can be parametrized as follows
\begin{eqnarray}
\label{ratios derivative}   \dfrac{\lim_{x\rightarrow\frac{\pi}{2}^+}\partial_x\psi^\alpha}{\lim_{x\rightarrow\frac{\pi}{2}^-}\partial_x\psi^\alpha}&=&\tanh\left(\frac{\pi}{2}\right)\dfrac{\cos{\left(\frac{\alpha}{2}\right)}+\sin\left(\frac{\alpha}{2}\right)}{\cos\left(\frac{\alpha}{2}\right)-\sin\left(\frac{\alpha}{2}\right)}\nonumber\\
&=&-\tan(\beta).
\end{eqnarray}
Thus, it is possible to translate information about $\alpha$ to $\beta$ as the label that characterizes the self-adjoint extensions, where $\beta\in[0,\pi)$, $\tan(\beta)\geq 0$, which is bijective in $U^\alpha$. This condition can be interpreted as a gluing condition at $b=+ b_0 $ and $b=\pi-b_0$ \cite{Pawlowski:2011zf,Assanioussi:2018hee,Assanioussi:2019iye}.\newline
To find the spectrum of the gravitational operator $ \Hat{\Theta}_{\lambda,g}$, we need to solve the eigenvalue problem for $\sigma=12\pi G k^2>0$ using eq.(\ref{KG op}). The eigenfunctions are given by \cite{Pawlowski:2011zf,Assanioussi:2018hee,Assanioussi:2019iye} 
\begin{eqnarray}
\label{eigen function}
    \psi_{k}(x)= A \begin{cases}\cosh \left(\frac{k\pi}{2}\right) e^{-ik\left(|x|-\frac{\pi}{2}\right)}, & |x|>\frac{\pi}{2}, \\ e^{k x}+e^{-kx},  & |x| \leq \frac{\pi}{2},\end{cases}
\end{eqnarray}
where $A\in \mathbb{C}$.\newline
Combining the boundary condition (\ref{ratios derivative}) with extended domain $D_\alpha$ of the self-adjoint extensions (\ref{extended domain}) the normalized eigenfunctions can be written as \cite{Pawlowski:2011zf,Assanioussi:2018hee,Assanioussi:2019iye}
\begin{eqnarray}
\label{normalized eigenfunction}
   \psi_{\beta, k}(x)=\zeta \begin{cases}\cos (k|x|+\varphi(\beta, k)), & |x|>\pi / 2, \\ \frac{\cos (k \pi / 2+\varphi(\beta, k))}{\cosh (k \pi / 2)} \cosh (k x), & |x| \leq \pi / 2,\end{cases}
\end{eqnarray}
where the constant of normalization is given by $\zeta$ and $\varphi(\beta, k)$ takes values given by the following transcendental equation 
\begin{equation}
    \label{transcendental eq}
    \tan (k \pi / 2+\varphi(\beta, k))=\tan (\beta) \tanh (k \pi / 2) .
\end{equation}

It turns out the spectrum of the gravitational operator, namely $k$, is continuous \cite{Assanioussi:2018hee,Assanioussi:2019iye}.

With the eigenfunctions (\ref{normalized eigenfunction}) it is possible now to ensure unitary evolution for $\lambda>0$. This is given in the following form 
\begin{equation}
    \psi_{\beta,k}(x,\phi)=e^{i(\phi-\phi_0)\sqrt{|\widehat{\Theta}_\beta|}}\psi_{\beta,k}(x,\phi_0),
\end{equation}
where the square root is defined for the positive part of the operator $\widehat{\Theta}_\beta$, eq. (\ref{sgn KG op}): $\psi_{\beta,k}(x,\phi)\in\mathcal{H}_{\beta}^{phy}=P_\beta^+\mathcal{H}_{kin}^{g}$, namely $P^+$ is the projection onto the positive part of $\widehat{\Theta}_\beta$.

We can see we have a continuous of possibilities parametrized self-adjoint  extensions in terms of $\beta$. This effect for eigenfunctions of $\widehat{\Theta}_\beta$ can be translated into generic state once we incorporate it into a propagator. That is what we do next. 

\section{Propagator}
\label{section:Eff model}
Evolution for states beyond eigenestates can be given in terms of the propagator as follows
\begin{equation}
\psi(y,\phi) = \int_{\mathbb{R}} \mathrm{d}x K_{\lambda}(y,\phi;x,\phi_0) \psi_0(x,\phi_0).
\end{equation}
The explicit form of $K_{\lambda}$ depends on the range of $\lambda$ as we can expect.

\textit{Case $\lambda\leq 0$}. Using the eigenfunctions (\ref{psi lneg0}) together with the generator of time evolution here being $\widehat{\Theta}_{\lambda\leq 0,g}$ we get
\begin{eqnarray}
    K_{\lambda\leq 0} &=&\int_{0}^{\infty} 2\sqrt{12\pi G}k\mathrm{d}k \ e^{-i \sqrt{\sigma(k)} (\phi-\phi_0)} \psi^*_{k}(y) \psi_{k}(x) \\
    &=& -i 32 (12 \pi G)^{3/2}(\phi-\phi_0) \left[\frac{1}{12 \pi G (\phi-\phi_0)^2-\left(x-y\right)^2}\right.\nonumber\\
    && +\left. \frac{1}{ 12 \pi G (\phi-\phi_0)^2-\left( x+y\right)^2}\right].
\end{eqnarray}

\textit{Case $\lambda> 0$}. Now we have to consider the eigenfunctions 
 (\ref{normalized eigenfunction}) and spectrum (\ref{transcendental eq}) of $\widehat{\Theta}_{\lambda>0,g}$, eq. (\ref{sgn KG op}). This yields
 \begin{eqnarray}
     K_{\lambda>0,\beta}=\int_{0}^{\infty}2\sqrt{12\pi G}k\mathrm{d}k  \ e^{-i \sqrt{\sigma(k)} (\phi-\phi_0)} \psi^*_{\beta,k}(y) \psi_{\beta,k}(x) \label{Kl>0}
 \end{eqnarray}
the subindex $\beta$ indicating the specific self-adjoint extension to be considered. Also, attention must be paid to the different possibilities depending on whether $|x|, |y| \lessgtr \frac{\pi}{2} $. We present here only the case $|x|, |y| > \frac{\pi}{2} $ but the others behave similarly. An explicit form can be obtained if in addition we use the large eigenvalue regime approximation allowing to use the following for (\ref{transcendental eq}) 
\begin{equation} 
\label{boundary condition asymptomatic }
    \varphi(\beta, k) =-\frac{k \pi}{2}(1-\tan (\beta))+O(e^{-k\pi}),
\end{equation}
using (\ref{boundary condition asymptomatic }) in (\ref{Kl>0}). This yields
\begin{eqnarray}
    \label{Green fucntion integral}
K_{\lambda>0,\beta}&&= -i 32(12\pi G)^{3/2}(\phi-\phi_0) \left[\frac{1}{12 \pi G (\phi-\phi_0)^2-\left(x-y\right)^2}\right.\nonumber\\
    && +\left. \frac{1}{ 12 \pi G (\phi-\phi_0)^2-\left[ x+y-\pi(1-\tan \beta)\right]^2}\right].
\end{eqnarray}
We can see that the self-adjoint extension corresponding to $\beta=\pi/4$ plays no role in the evolution. On the other hand any  extension with $\beta\in (0,\pi/4)\cup(\pi/4,\pi]$ affects it. Semiclassical states would be rather interesting to analyze to understand better the effect of the extensions. We leave this problem for future work.

\section{Discussion and Conclusions}
\label{section:Conclusion}
This paper studies self-adjoint extensions of a generalized soluble LQC model that includes an arbitrary weight parameter between the Euclidean and the Lorentzian terms in the Hamiltonian constraint, thus generalizing previous work  \cite{Zhang:2021zfp,Yang:2009fp} but including them as particular cases  \cite{Pawlowski:2011zf,Assanioussi:2018hee,Assanioussi:2019iye,Yang:2009fp,Zhang:2021zfp,Ashtekar:2006wn,Ashtekar:2007em, Kaminski:2007gm, Kaminski:2009pc}. The inclusion of the Lorentzian term is relevant because in effective models it gives rise to an emergent cosmological constant. First unitary evolution was ensured by considering self-adjoint extensions of the Hamiltonian constraint with a specific wight value corresponding to classical GR but the effective cosmological constant turned out to be Planckian \cite{Pawlowski:2011zf,Assanioussi:2018hee,Assanioussi:2019iye}. Then a weight parameter was determined to make the effective cosmological constant equal to the observed one \cite{Zhang:2021zfp}. However, unitary evolution was not considered.

Following \cite{Assanioussi:2018hee,Assanioussi:2019iye}, we have characterized the unitary evolution of the model for all real values of the weight parameter in the Hamiltonian constraint. For $\lambda\leq 0$ the Hamiltonian constraint is essentially self-adjoint and unitary evolution follows, whilst for $\lambda>0$ self-adjoint extensions are need to get it. Previous results on unitary evolution are contained in ours for different values of the weight parameter.  $\lambda=-1/\gamma^2$ yields the purely Euclidean term \cite{Kaminski:2007gm,Kaminski:2009pc} and $\lambda=0$ is just the purely Lorentzian term , in both cases, the Hamiltonian constraint is essentially self-adjoint. $\lambda=1$, on the other hand, is the weight corresponding to classical GR and it requires self-adjoint extensions to evolve unitarily
\cite{Assanioussi:2018hee,Assanioussi:2019iye}. Now for the case in \cite{Zhang:2021zfp} with $\lambda_0\sim 10^{-122}$ tiny but positive,  self-adjoint extensions are mandatory to get physical (unitary) evolution.

Our study made used of von Neumann's deficit indices technique. Firstly we identified the unitary evolution of eigenstates of the Hamiltonian constraint in both cases $\lambda\leq 0$, for which it is essentially self-adjoint, and $\lambda>0$, where self-adjoint extensions are required. For general states we determined a propagator for both cases showing the specific effect of the extensions in the second.

As for future work we can consider the following.  We can generalize our study for the inclusion of an explicit cosmological constant such as in \cite{Kaminski:2009pc,Pawlowski:2011zf} as well as a weight parameter in the Hamiltonian constraint. In this case the  Hamiltonian constraint is expected to need self-adjoint extensions yielding unitary evolution parametrized by the group  $U(2)$ rather than $U(1)$ of the present work. We adopted for simplicity the soluble model LQC which is flat so further work is necessary to include spatial curvature.
Finally, to disentangle the physical implications of the extensions for positive weight parameter that include the cases of $\lambda=\lambda_0,1$ it is desirable to get an effective theory along the lines of semiclassical techniques \cite{Ashtekar:2010gz, Yang:2009fp,Zhang:2021zfp, Flores-Gonzalez:2013zuk,Qin:2012gaa,Huang:2011es}.

\color{black}

\section{Acknowledgments}
\label{section:Acknowledgments}
O.G would like to thank Prof. T. Pawłowski for discussions on self-adjointness in LQC and hospitality during a visit to University of Wrocław. Partial support is acknowledged from CONAHCyT M\'exico under grants: A1-S-8742, 304001, 376127; Project No. 269652 and Fronteras Project 281; I0101/131/07 C-234/07 of the collaboration of the Instituto Avanzado de Cosmolog\'ia (IAC) collaboration (http:// www.iac.edu.mx). O.G. thanks that this work was supported by UNAM Posdoctoral Program (POSDOC). The work of T.M. was partially supported by CONAHCyT through the Fondo Sectorial de Investigaci\'on para la Educaci\'on, grant CB-2014-1, No. 240512. Partial support is acknowledged also from CONAHCyT-México under grant CBF-2023-2024-1937. HAMT was partially supported by mexican Sistema Nacional de Investigadoras e Investigadores SNII 14585.

\bibliography{ref}% Produces the bibliography via BibTeX.

\begin{thebibliography}{39}%
\makeatletter
\providecommand \@ifxundefined [1]{%
 \@ifx{#1\undefined}
}%
\providecommand \@ifnum [1]{%
 \ifnum #1\expandafter \@firstoftwo
 \else \expandafter \@secondoftwo
 \fi
}%
\providecommand \@ifx [1]{%
 \ifx #1\expandafter \@firstoftwo
 \else \expandafter \@secondoftwo
 \fi
}%
\providecommand \natexlab [1]{#1}%
\providecommand \enquote  [1]{``#1''}%
\providecommand \bibnamefont  [1]{#1}%
\providecommand \bibfnamefont [1]{#1}%
\providecommand \citenamefont [1]{#1}%
\providecommand \href@noop [0]{\@secondoftwo}%
\providecommand \href [0]{\begingroup \@sanitize@url \@href}%
\providecommand \@href[1]{\@@startlink{#1}\@@href}%
\providecommand \@@href[1]{\endgroup#1\@@endlink}%
\providecommand \@sanitize@url [0]{\catcode `\\12\catcode `\$12\catcode
  `\&12\catcode `\#12\catcode `\^12\catcode `\_12\catcode `\%12\relax}%
\providecommand \@@startlink[1]{}%
\providecommand \@@endlink[0]{}%
\providecommand \url  [0]{\begingroup\@sanitize@url \@url }%
\providecommand \@url [1]{\endgroup\@href {#1}{\urlprefix }}%
\providecommand \urlprefix  [0]{URL }%
\providecommand \Eprint [0]{\href }%
\providecommand \doibase [0]{https://doi.org/}%
\providecommand \selectlanguage [0]{\@gobble}%
\providecommand \bibinfo  [0]{\@secondoftwo}%
\providecommand \bibfield  [0]{\@secondoftwo}%
\providecommand \translation [1]{[#1]}%
\providecommand \BibitemOpen [0]{}%
\providecommand \bibitemStop [0]{}%
\providecommand \bibitemNoStop [0]{.\EOS\space}%
\providecommand \EOS [0]{\spacefactor3000\relax}%
\providecommand \BibitemShut  [1]{\csname bibitem#1\endcsname}%
\let\auto@bib@innerbib\@empty
%</preamble>
\bibitem [{\citenamefont {Aghanim}\ \emph
  {et~al.}(2020{\natexlab{a}})\citenamefont {Aghanim} \emph
  {et~al.}}]{Planck:2018nkj}%
  \BibitemOpen
  \bibfield  {author} {\bibinfo {author} {\bibfnamefont {N.}~\bibnamefont
  {Aghanim}} \emph {et~al.} (\bibinfo {collaboration} {Planck}),\ }\bibfield
  {title} {\bibinfo {title} {{Planck 2018 results. I. Overview and the
  cosmological legacy of Planck}},\ }\href
  {https://doi.org/10.1051/0004-6361/201833880} {\bibfield  {journal} {\bibinfo
   {journal} {Astron. Astrophys.}\ }\textbf {\bibinfo {volume} {641}},\
  \bibinfo {pages} {A1} (\bibinfo {year} {2020}{\natexlab{a}})},\ \Eprint
  {https://arxiv.org/abs/1807.06205} {arXiv:1807.06205 [astro-ph.CO]}
  \BibitemShut {NoStop}%
\bibitem [{\citenamefont {Aghanim}\ \emph
  {et~al.}(2020{\natexlab{b}})\citenamefont {Aghanim} \emph
  {et~al.}}]{Planck:2018vyg}%
  \BibitemOpen
  \bibfield  {author} {\bibinfo {author} {\bibfnamefont {N.}~\bibnamefont
  {Aghanim}} \emph {et~al.} (\bibinfo {collaboration} {Planck}),\ }\bibfield
  {title} {\bibinfo {title} {{Planck 2018 results. VI. Cosmological
  parameters}},\ }\href {https://doi.org/10.1051/0004-6361/201833910}
  {\bibfield  {journal} {\bibinfo  {journal} {Astron. Astrophys.}\ }\textbf
  {\bibinfo {volume} {641}},\ \bibinfo {pages} {A6} (\bibinfo {year}
  {2020}{\natexlab{b}})},\ \bibinfo {note} {[Erratum: Astron.Astrophys. 652, C4
  (2021)]},\ \Eprint {https://arxiv.org/abs/1807.06209} {arXiv:1807.06209
  [astro-ph.CO]} \BibitemShut {NoStop}%
\bibitem [{\citenamefont {Adame}\ \emph {et~al.}(2024)\citenamefont {Adame}
  \emph {et~al.}}]{DESI:2024mwx}%
  \BibitemOpen
  \bibfield  {author} {\bibinfo {author} {\bibfnamefont {A.~G.}\ \bibnamefont
  {Adame}} \emph {et~al.} (\bibinfo {collaboration} {DESI}),\ }\bibfield
  {title} {\bibinfo {title} {{DESI 2024 VI: Cosmological Constraints from the
  Measurements of Baryon Acoustic Oscillations}},\ }\href@noop {} {\  (\bibinfo
  {year} {2024})},\ \Eprint {https://arxiv.org/abs/2404.03002}
  {arXiv:2404.03002 [astro-ph.CO]} \BibitemShut {NoStop}%
\bibitem [{\citenamefont {Weinberg}(1989)}]{Weinberg:1988cp}%
  \BibitemOpen
  \bibfield  {author} {\bibinfo {author} {\bibfnamefont {S.}~\bibnamefont
  {Weinberg}},\ }\bibfield  {title} {\bibinfo {title} {{The Cosmological
  Constant Problem}},\ }\href {https://doi.org/10.1103/RevModPhys.61.1}
  {\bibfield  {journal} {\bibinfo  {journal} {Rev. Mod. Phys.}\ }\textbf
  {\bibinfo {volume} {61}},\ \bibinfo {pages} {1} (\bibinfo {year}
  {1989})}\BibitemShut {NoStop}%
\bibitem [{\citenamefont {Rovelli}(2004)}]{Rovelli:2004tv}%
  \BibitemOpen
  \bibfield  {author} {\bibinfo {author} {\bibfnamefont {C.}~\bibnamefont
  {Rovelli}},\ }\href {https://doi.org/10.1017/CBO9780511755804} {\emph
  {\bibinfo {title} {{Quantum gravity}}}},\ Cambridge Monographs on
  Mathematical Physics\ (\bibinfo  {publisher} {Univ. Pr.},\ \bibinfo {address}
  {Cambridge, UK},\ \bibinfo {year} {2004})\BibitemShut {NoStop}%
\bibitem [{\citenamefont {Thiemann}(2007)}]{Thiemann:2007pyv}%
  \BibitemOpen
  \bibfield  {author} {\bibinfo {author} {\bibfnamefont {T.}~\bibnamefont
  {Thiemann}},\ }\href {https://doi.org/10.1017/CBO9780511755682} {\emph
  {\bibinfo {title} {{Modern Canonical Quantum General Relativity}}}},\
  Cambridge Monographs on Mathematical Physics\ (\bibinfo  {publisher}
  {Cambridge University Press},\ \bibinfo {year} {2007})\BibitemShut {NoStop}%
\bibitem [{\citenamefont {Morales-T\'ecotl}\ \emph {et~al.}(2021)\citenamefont
  {Morales-T\'ecotl}, \citenamefont {Rastgoo},\ and\ \citenamefont
  {Ruelas}}]{Morales-Tecotl:2018ugi}%
  \BibitemOpen
  \bibfield  {author} {\bibinfo {author} {\bibfnamefont {H.~A.}\ \bibnamefont
  {Morales-T\'ecotl}}, \bibinfo {author} {\bibfnamefont {S.}~\bibnamefont
  {Rastgoo}},\ and\ \bibinfo {author} {\bibfnamefont {J.~C.}\ \bibnamefont
  {Ruelas}},\ }\bibfield  {title} {\bibinfo {title} {{Effective dynamics of the
  Schwarzschild black hole interior with inverse triad corrections}},\ }\href
  {https://doi.org/10.1016/j.aop.2021.168401} {\bibfield  {journal} {\bibinfo
  {journal} {Annals Phys.}\ }\textbf {\bibinfo {volume} {426}},\ \bibinfo
  {pages} {168401} (\bibinfo {year} {2021})},\ \Eprint
  {https://arxiv.org/abs/1806.05795} {arXiv:1806.05795 [gr-qc]} \BibitemShut
  {NoStop}%
\bibitem [{\citenamefont {Bojowald}(2001)}]{Bojowald:2001xe}%
  \BibitemOpen
  \bibfield  {author} {\bibinfo {author} {\bibfnamefont {M.}~\bibnamefont
  {Bojowald}},\ }\bibfield  {title} {\bibinfo {title} {{Absence of singularity
  in loop quantum cosmology}},\ }\href
  {https://doi.org/10.1103/PhysRevLett.86.5227} {\bibfield  {journal} {\bibinfo
   {journal} {Phys. Rev. Lett.}\ }\textbf {\bibinfo {volume} {86}},\ \bibinfo
  {pages} {5227} (\bibinfo {year} {2001})},\ \Eprint
  {https://arxiv.org/abs/gr-qc/0102069} {arXiv:gr-qc/0102069} \BibitemShut
  {NoStop}%
\bibitem [{\citenamefont {Bojowald}(2005)}]{Bojowald:2005epg}%
  \BibitemOpen
  \bibfield  {author} {\bibinfo {author} {\bibfnamefont {M.}~\bibnamefont
  {Bojowald}},\ }\bibfield  {title} {\bibinfo {title} {{Loop quantum
  cosmology}},\ }\href {https://doi.org/10.12942/lrr-2005-11} {\bibfield
  {journal} {\bibinfo  {journal} {Living Rev. Rel.}\ }\textbf {\bibinfo
  {volume} {8}},\ \bibinfo {pages} {11} (\bibinfo {year} {2005})},\ \Eprint
  {https://arxiv.org/abs/gr-qc/0601085} {arXiv:gr-qc/0601085} \BibitemShut
  {NoStop}%
\bibitem [{\citenamefont {Ashtekar}\ \emph {et~al.}(2003)\citenamefont
  {Ashtekar}, \citenamefont {Bojowald},\ and\ \citenamefont
  {Lewandowski}}]{Ashtekar:2003hd}%
  \BibitemOpen
  \bibfield  {author} {\bibinfo {author} {\bibfnamefont {A.}~\bibnamefont
  {Ashtekar}}, \bibinfo {author} {\bibfnamefont {M.}~\bibnamefont {Bojowald}},\
  and\ \bibinfo {author} {\bibfnamefont {J.}~\bibnamefont {Lewandowski}},\
  }\bibfield  {title} {\bibinfo {title} {{Mathematical structure of loop
  quantum cosmology}},\ }\href {https://doi.org/10.4310/ATMP.2003.v7.n2.a2}
  {\bibfield  {journal} {\bibinfo  {journal} {Adv. Theor. Math. Phys.}\
  }\textbf {\bibinfo {volume} {7}},\ \bibinfo {pages} {233} (\bibinfo {year}
  {2003})},\ \Eprint {https://arxiv.org/abs/gr-qc/0304074}
  {arXiv:gr-qc/0304074} \BibitemShut {NoStop}%
\bibitem [{\citenamefont {Ashtekar}\ \emph {et~al.}(2007)\citenamefont
  {Ashtekar}, \citenamefont {Pawlowski}, \citenamefont {Singh},\ and\
  \citenamefont {Vandersloot}}]{Ashtekar:2006es}%
  \BibitemOpen
  \bibfield  {author} {\bibinfo {author} {\bibfnamefont {A.}~\bibnamefont
  {Ashtekar}}, \bibinfo {author} {\bibfnamefont {T.}~\bibnamefont {Pawlowski}},
  \bibinfo {author} {\bibfnamefont {P.}~\bibnamefont {Singh}},\ and\ \bibinfo
  {author} {\bibfnamefont {K.}~\bibnamefont {Vandersloot}},\ }\bibfield
  {title} {\bibinfo {title} {{Loop quantum cosmology of k=1 FRW models}},\
  }\href {https://doi.org/10.1103/PhysRevD.75.024035} {\bibfield  {journal}
  {\bibinfo  {journal} {Phys. Rev. D}\ }\textbf {\bibinfo {volume} {75}},\
  \bibinfo {pages} {024035} (\bibinfo {year} {2007})},\ \Eprint
  {https://arxiv.org/abs/gr-qc/0612104} {arXiv:gr-qc/0612104} \BibitemShut
  {NoStop}%
\bibitem [{\citenamefont {Ashtekar}\ \emph {et~al.}(2006)\citenamefont
  {Ashtekar}, \citenamefont {Pawlowski},\ and\ \citenamefont
  {Singh}}]{Ashtekar:2006wn}%
  \BibitemOpen
  \bibfield  {author} {\bibinfo {author} {\bibfnamefont {A.}~\bibnamefont
  {Ashtekar}}, \bibinfo {author} {\bibfnamefont {T.}~\bibnamefont
  {Pawlowski}},\ and\ \bibinfo {author} {\bibfnamefont {P.}~\bibnamefont
  {Singh}},\ }\bibfield  {title} {\bibinfo {title} {{Quantum Nature of the Big
  Bang: Improved dynamics}},\ }\href
  {https://doi.org/10.1103/PhysRevD.74.084003} {\bibfield  {journal} {\bibinfo
  {journal} {Phys. Rev. D}\ }\textbf {\bibinfo {volume} {74}},\ \bibinfo
  {pages} {084003} (\bibinfo {year} {2006})},\ \Eprint
  {https://arxiv.org/abs/gr-qc/0607039} {arXiv:gr-qc/0607039} \BibitemShut
  {NoStop}%
\bibitem [{\citenamefont {Pawlowski}\ and\ \citenamefont
  {Ashtekar}(2012)}]{Pawlowski:2011zf}%
  \BibitemOpen
  \bibfield  {author} {\bibinfo {author} {\bibfnamefont {T.}~\bibnamefont
  {Pawlowski}}\ and\ \bibinfo {author} {\bibfnamefont {A.}~\bibnamefont
  {Ashtekar}},\ }\bibfield  {title} {\bibinfo {title} {{Positive cosmological
  constant in loop quantum cosmology}},\ }\href
  {https://doi.org/10.1103/PhysRevD.85.064001} {\bibfield  {journal} {\bibinfo
  {journal} {Phys. Rev. D}\ }\textbf {\bibinfo {volume} {85}},\ \bibinfo
  {pages} {064001} (\bibinfo {year} {2012})},\ \Eprint
  {https://arxiv.org/abs/1112.0360} {arXiv:1112.0360 [gr-qc]} \BibitemShut
  {NoStop}%
\bibitem [{\citenamefont {Kaminski}\ and\ \citenamefont
  {Lewandowski}(2008)}]{Kaminski:2007gm}%
  \BibitemOpen
  \bibfield  {author} {\bibinfo {author} {\bibfnamefont {W.}~\bibnamefont
  {Kaminski}}\ and\ \bibinfo {author} {\bibfnamefont {J.}~\bibnamefont
  {Lewandowski}},\ }\bibfield  {title} {\bibinfo {title} {{The Flat FRW model
  in LQC: The Self-adjointness}},\ }\href
  {https://doi.org/10.1088/0264-9381/25/3/035001} {\bibfield  {journal}
  {\bibinfo  {journal} {Class. Quant. Grav.}\ }\textbf {\bibinfo {volume}
  {25}},\ \bibinfo {pages} {035001} (\bibinfo {year} {2008})},\ \Eprint
  {https://arxiv.org/abs/0709.3120} {arXiv:0709.3120 [gr-qc]} \BibitemShut
  {NoStop}%
\bibitem [{\citenamefont {Kaminski}\ and\ \citenamefont
  {Pawlowski}(2010)}]{Kaminski:2009pc}%
  \BibitemOpen
  \bibfield  {author} {\bibinfo {author} {\bibfnamefont {W.}~\bibnamefont
  {Kaminski}}\ and\ \bibinfo {author} {\bibfnamefont {T.}~\bibnamefont
  {Pawlowski}},\ }\bibfield  {title} {\bibinfo {title} {{The LQC evolution
  operator of FRW universe with positive cosmological constant}},\ }\href
  {https://doi.org/10.1103/PhysRevD.81.024014} {\bibfield  {journal} {\bibinfo
  {journal} {Phys. Rev. D}\ }\textbf {\bibinfo {volume} {81}},\ \bibinfo
  {pages} {024014} (\bibinfo {year} {2010})},\ \Eprint
  {https://arxiv.org/abs/0912.0162} {arXiv:0912.0162 [gr-qc]} \BibitemShut
  {NoStop}%
\bibitem [{\citenamefont {Bentivegna}\ and\ \citenamefont
  {Pawlowski}(2008)}]{Bentivegna:2008bg}%
  \BibitemOpen
  \bibfield  {author} {\bibinfo {author} {\bibfnamefont {E.}~\bibnamefont
  {Bentivegna}}\ and\ \bibinfo {author} {\bibfnamefont {T.}~\bibnamefont
  {Pawlowski}},\ }\bibfield  {title} {\bibinfo {title} {{Anti-deSitter universe
  dynamics in LQC}},\ }\href {https://doi.org/10.1103/PhysRevD.77.124025}
  {\bibfield  {journal} {\bibinfo  {journal} {Phys. Rev. D}\ }\textbf {\bibinfo
  {volume} {77}},\ \bibinfo {pages} {124025} (\bibinfo {year} {2008})},\
  \Eprint {https://arxiv.org/abs/0803.4446} {arXiv:0803.4446 [gr-qc]}
  \BibitemShut {NoStop}%
\bibitem [{\citenamefont {Martin-Benito}\ \emph {et~al.}(2009)\citenamefont
  {Martin-Benito}, \citenamefont {Marugan},\ and\ \citenamefont
  {Olmedo}}]{Martin-Benito:2009htq}%
  \BibitemOpen
  \bibfield  {author} {\bibinfo {author} {\bibfnamefont {M.}~\bibnamefont
  {Martin-Benito}}, \bibinfo {author} {\bibfnamefont {G.~A.~M.}\ \bibnamefont
  {Marugan}},\ and\ \bibinfo {author} {\bibfnamefont {J.}~\bibnamefont
  {Olmedo}},\ }\bibfield  {title} {\bibinfo {title} {{Further Improvements in
  the Understanding of Isotropic Loop Quantum Cosmology}},\ }\href
  {https://doi.org/10.1103/PhysRevD.80.104015} {\bibfield  {journal} {\bibinfo
  {journal} {Phys. Rev. D}\ }\textbf {\bibinfo {volume} {80}},\ \bibinfo
  {pages} {104015} (\bibinfo {year} {2009})},\ \Eprint
  {https://arxiv.org/abs/0909.2829} {arXiv:0909.2829 [gr-qc]} \BibitemShut
  {NoStop}%
\bibitem [{\citenamefont {Ashtekar}\ \emph {et~al.}(2008)\citenamefont
  {Ashtekar}, \citenamefont {Corichi},\ and\ \citenamefont
  {Singh}}]{Ashtekar:2007em}%
  \BibitemOpen
  \bibfield  {author} {\bibinfo {author} {\bibfnamefont {A.}~\bibnamefont
  {Ashtekar}}, \bibinfo {author} {\bibfnamefont {A.}~\bibnamefont {Corichi}},\
  and\ \bibinfo {author} {\bibfnamefont {P.}~\bibnamefont {Singh}},\ }\bibfield
   {title} {\bibinfo {title} {{Robustness of key features of loop quantum
  cosmology}},\ }\href {https://doi.org/10.1103/PhysRevD.77.024046} {\bibfield
  {journal} {\bibinfo  {journal} {Phys. Rev. D}\ }\textbf {\bibinfo {volume}
  {77}},\ \bibinfo {pages} {024046} (\bibinfo {year} {2008})},\ \Eprint
  {https://arxiv.org/abs/0710.3565} {arXiv:0710.3565 [gr-qc]} \BibitemShut
  {NoStop}%
\bibitem [{\citenamefont {Szulc}\ \emph {et~al.}(2007)\citenamefont {Szulc},
  \citenamefont {Kaminski},\ and\ \citenamefont {Lewandowski}}]{Szulc:2006ep}%
  \BibitemOpen
  \bibfield  {author} {\bibinfo {author} {\bibfnamefont {L.}~\bibnamefont
  {Szulc}}, \bibinfo {author} {\bibfnamefont {W.}~\bibnamefont {Kaminski}},\
  and\ \bibinfo {author} {\bibfnamefont {J.}~\bibnamefont {Lewandowski}},\
  }\bibfield  {title} {\bibinfo {title} {{Closed FRW model in Loop Quantum
  Cosmology}},\ }\href {https://doi.org/10.1088/0264-9381/24/10/008} {\bibfield
   {journal} {\bibinfo  {journal} {Class. Quant. Grav.}\ }\textbf {\bibinfo
  {volume} {24}},\ \bibinfo {pages} {2621} (\bibinfo {year} {2007})},\ \Eprint
  {https://arxiv.org/abs/gr-qc/0612101} {arXiv:gr-qc/0612101} \BibitemShut
  {NoStop}%
\bibitem [{\citenamefont {Szulc}(2007)}]{Szulc:2007uk}%
  \BibitemOpen
  \bibfield  {author} {\bibinfo {author} {\bibfnamefont {L.}~\bibnamefont
  {Szulc}},\ }\bibfield  {title} {\bibinfo {title} {{Open FRW model in Loop
  Quantum Cosmology}},\ }\href {https://doi.org/10.1088/0264-9381/24/24/003}
  {\bibfield  {journal} {\bibinfo  {journal} {Class. Quant. Grav.}\ }\textbf
  {\bibinfo {volume} {24}},\ \bibinfo {pages} {6191} (\bibinfo {year}
  {2007})},\ \Eprint {https://arxiv.org/abs/0707.1816} {arXiv:0707.1816
  [gr-qc]} \BibitemShut {NoStop}%
\bibitem [{\citenamefont {Yang}\ \emph {et~al.}(2009)\citenamefont {Yang},
  \citenamefont {Ding},\ and\ \citenamefont {Ma}}]{Yang:2009fp}%
  \BibitemOpen
  \bibfield  {author} {\bibinfo {author} {\bibfnamefont {J.}~\bibnamefont
  {Yang}}, \bibinfo {author} {\bibfnamefont {Y.}~\bibnamefont {Ding}},\ and\
  \bibinfo {author} {\bibfnamefont {Y.}~\bibnamefont {Ma}},\ }\bibfield
  {title} {\bibinfo {title} {{Alternative quantization of the Hamiltonian in
  loop quantum cosmology II: Including the Lorentz term}},\ }\href
  {https://doi.org/10.1016/j.physletb.2009.10.072} {\bibfield  {journal}
  {\bibinfo  {journal} {Phys. Lett. B}\ }\textbf {\bibinfo {volume} {682}},\
  \bibinfo {pages} {1} (\bibinfo {year} {2009})},\ \Eprint
  {https://arxiv.org/abs/0904.4379} {arXiv:0904.4379 [gr-qc]} \BibitemShut
  {NoStop}%
\bibitem [{\citenamefont {Zhang}\ \emph {et~al.}(2021)\citenamefont {Zhang},
  \citenamefont {Long},\ and\ \citenamefont {Ma}}]{Zhang:2021zfp}%
  \BibitemOpen
  \bibfield  {author} {\bibinfo {author} {\bibfnamefont {X.}~\bibnamefont
  {Zhang}}, \bibinfo {author} {\bibfnamefont {G.}~\bibnamefont {Long}},\ and\
  \bibinfo {author} {\bibfnamefont {Y.}~\bibnamefont {Ma}},\ }\bibfield
  {title} {\bibinfo {title} {{Loop quantum gravity and cosmological
  constant}},\ }\href {https://doi.org/10.1016/j.physletb.2021.136770}
  {\bibfield  {journal} {\bibinfo  {journal} {Phys. Lett. B}\ }\textbf
  {\bibinfo {volume} {823}},\ \bibinfo {pages} {136770} (\bibinfo {year}
  {2021})},\ \Eprint {https://arxiv.org/abs/2101.07527} {arXiv:2101.07527
  [gr-qc]} \BibitemShut {NoStop}%
\bibitem [{\citenamefont {Assanioussi}\ \emph {et~al.}(2018)\citenamefont
  {Assanioussi}, \citenamefont {Dapor}, \citenamefont {Liegener},\ and\
  \citenamefont {Paw\l{}owski}}]{Assanioussi:2018hee}%
  \BibitemOpen
  \bibfield  {author} {\bibinfo {author} {\bibfnamefont {M.}~\bibnamefont
  {Assanioussi}}, \bibinfo {author} {\bibfnamefont {A.}~\bibnamefont {Dapor}},
  \bibinfo {author} {\bibfnamefont {K.}~\bibnamefont {Liegener}},\ and\
  \bibinfo {author} {\bibfnamefont {T.}~\bibnamefont {Paw\l{}owski}},\
  }\bibfield  {title} {\bibinfo {title} {{Emergent de Sitter Epoch of the
  Quantum Cosmos from Loop Quantum Cosmology}},\ }\href
  {https://doi.org/10.1103/PhysRevLett.121.081303} {\bibfield  {journal}
  {\bibinfo  {journal} {Phys. Rev. Lett.}\ }\textbf {\bibinfo {volume} {121}},\
  \bibinfo {pages} {081303} (\bibinfo {year} {2018})},\ \Eprint
  {https://arxiv.org/abs/1801.00768} {arXiv:1801.00768 [gr-qc]} \BibitemShut
  {NoStop}%
\bibitem [{\citenamefont {Assanioussi}\ \emph {et~al.}(2019)\citenamefont
  {Assanioussi}, \citenamefont {Dapor}, \citenamefont {Liegener},\ and\
  \citenamefont {Paw\l{}owski}}]{Assanioussi:2019iye}%
  \BibitemOpen
  \bibfield  {author} {\bibinfo {author} {\bibfnamefont {M.}~\bibnamefont
  {Assanioussi}}, \bibinfo {author} {\bibfnamefont {A.}~\bibnamefont {Dapor}},
  \bibinfo {author} {\bibfnamefont {K.}~\bibnamefont {Liegener}},\ and\
  \bibinfo {author} {\bibfnamefont {T.}~\bibnamefont {Paw\l{}owski}},\
  }\bibfield  {title} {\bibinfo {title} {{Emergent de Sitter epoch of the Loop
  Quantum Cosmos: a detailed analysis}},\ }\href
  {https://doi.org/10.1103/PhysRevD.100.084003} {\bibfield  {journal} {\bibinfo
   {journal} {Phys. Rev. D}\ }\textbf {\bibinfo {volume} {100}},\ \bibinfo
  {pages} {084003} (\bibinfo {year} {2019})},\ \Eprint
  {https://arxiv.org/abs/1906.05315} {arXiv:1906.05315 [gr-qc]} \BibitemShut
  {NoStop}%
\bibitem [{\citenamefont {Yang}\ \emph {et~al.}(2023)\citenamefont {Yang},
  \citenamefont {Zhang},\ and\ \citenamefont {Zhang}}]{Yang:2022aec}%
  \BibitemOpen
  \bibfield  {author} {\bibinfo {author} {\bibfnamefont {J.}~\bibnamefont
  {Yang}}, \bibinfo {author} {\bibfnamefont {C.}~\bibnamefont {Zhang}},\ and\
  \bibinfo {author} {\bibfnamefont {X.}~\bibnamefont {Zhang}},\ }\bibfield
  {title} {\bibinfo {title} {{Alternative k=-1 loop quantum cosmology}},\
  }\href {https://doi.org/10.1103/PhysRevD.107.046012} {\bibfield  {journal}
  {\bibinfo  {journal} {Phys. Rev. D}\ }\textbf {\bibinfo {volume} {107}},\
  \bibinfo {pages} {046012} (\bibinfo {year} {2023})},\ \Eprint
  {https://arxiv.org/abs/2212.05748} {arXiv:2212.05748 [gr-qc]} \BibitemShut
  {NoStop}%
\bibitem [{\citenamefont {Meissner}(2004)}]{Meissner:2004ju}%
  \BibitemOpen
  \bibfield  {author} {\bibinfo {author} {\bibfnamefont {K.~A.}\ \bibnamefont
  {Meissner}},\ }\bibfield  {title} {\bibinfo {title} {{Black hole entropy in
  loop quantum gravity}},\ }\href {https://doi.org/10.1088/0264-9381/21/22/015}
  {\bibfield  {journal} {\bibinfo  {journal} {Class. Quant. Grav.}\ }\textbf
  {\bibinfo {volume} {21}},\ \bibinfo {pages} {5245} (\bibinfo {year}
  {2004})},\ \Eprint {https://arxiv.org/abs/gr-qc/0407052}
  {arXiv:gr-qc/0407052} \BibitemShut {NoStop}%
\bibitem [{\citenamefont {Ghosh}\ and\ \citenamefont
  {Mitra}(2005)}]{Ghosh:2004rq}%
  \BibitemOpen
  \bibfield  {author} {\bibinfo {author} {\bibfnamefont {A.}~\bibnamefont
  {Ghosh}}\ and\ \bibinfo {author} {\bibfnamefont {P.}~\bibnamefont {Mitra}},\
  }\bibfield  {title} {\bibinfo {title} {{A Bound on the log correction to the
  black hole area law}},\ }\href {https://doi.org/10.1103/PhysRevD.71.027502}
  {\bibfield  {journal} {\bibinfo  {journal} {Phys. Rev. D}\ }\textbf {\bibinfo
  {volume} {71}},\ \bibinfo {pages} {027502} (\bibinfo {year} {2005})},\
  \Eprint {https://arxiv.org/abs/gr-qc/0401070} {arXiv:gr-qc/0401070}
  \BibitemShut {NoStop}%
\bibitem [{\citenamefont {Rovelli}(1996)}]{Rovelli:1996dv}%
  \BibitemOpen
  \bibfield  {author} {\bibinfo {author} {\bibfnamefont {C.}~\bibnamefont
  {Rovelli}},\ }\bibfield  {title} {\bibinfo {title} {{Black hole entropy from
  loop quantum gravity}},\ }\href {https://doi.org/10.1103/PhysRevLett.77.3288}
  {\bibfield  {journal} {\bibinfo  {journal} {Phys. Rev. Lett.}\ }\textbf
  {\bibinfo {volume} {77}},\ \bibinfo {pages} {3288} (\bibinfo {year}
  {1996})},\ \Eprint {https://arxiv.org/abs/gr-qc/9603063}
  {arXiv:gr-qc/9603063} \BibitemShut {NoStop}%
\bibitem [{\citenamefont {Ashtekar}\ \emph {et~al.}(1998)\citenamefont
  {Ashtekar}, \citenamefont {Baez}, \citenamefont {Corichi},\ and\
  \citenamefont {Krasnov}}]{Ashtekar:1997yu}%
  \BibitemOpen
  \bibfield  {author} {\bibinfo {author} {\bibfnamefont {A.}~\bibnamefont
  {Ashtekar}}, \bibinfo {author} {\bibfnamefont {J.}~\bibnamefont {Baez}},
  \bibinfo {author} {\bibfnamefont {A.}~\bibnamefont {Corichi}},\ and\ \bibinfo
  {author} {\bibfnamefont {K.}~\bibnamefont {Krasnov}},\ }\bibfield  {title}
  {\bibinfo {title} {{Quantum geometry and black hole entropy}},\ }\href
  {https://doi.org/10.1103/PhysRevLett.80.904} {\bibfield  {journal} {\bibinfo
  {journal} {Phys. Rev. Lett.}\ }\textbf {\bibinfo {volume} {80}},\ \bibinfo
  {pages} {904} (\bibinfo {year} {1998})},\ \Eprint
  {https://arxiv.org/abs/gr-qc/9710007} {arXiv:gr-qc/9710007} \BibitemShut
  {NoStop}%
\bibitem [{\citenamefont {Reed}\ and\ \citenamefont
  {Simon}(1972)}]{Reed:1972uy}%
  \BibitemOpen
  \bibfield  {author} {\bibinfo {author} {\bibfnamefont {M.}~\bibnamefont
  {Reed}}\ and\ \bibinfo {author} {\bibfnamefont {B.}~\bibnamefont {Simon}},\
  }\href@noop {} {\emph {\bibinfo {title} {{Methods of Modern Mathematical
  Physics. I. Functional analysis}}}}\ (\bibinfo {year} {1972})\BibitemShut
  {NoStop}%
\bibitem [{\citenamefont {Reed}\ and\ \citenamefont
  {Simon}(1975)}]{Reed:1975uy}%
  \BibitemOpen
  \bibfield  {author} {\bibinfo {author} {\bibfnamefont {M.}~\bibnamefont
  {Reed}}\ and\ \bibinfo {author} {\bibfnamefont {B.}~\bibnamefont {Simon}},\
  }\href@noop {} {\emph {\bibinfo {title} {{Methods of Modern Mathematical
  Physics. 2. Fourier Analysis, Self-adjointness}}}}\ (\bibinfo {year}
  {1975})\BibitemShut {NoStop}%
\bibitem [{\citenamefont {Simon}(1998)}]{Simon98theclassical}%
  \BibitemOpen
  \bibfield  {author} {\bibinfo {author} {\bibfnamefont {B.}~\bibnamefont
  {Simon}},\ }\bibfield  {title} {\bibinfo {title} {The classical moment
  problem as a self-adjoint finite difference operator},\ }\href@noop {}
  {\bibfield  {journal} {\bibinfo  {journal} {Adv. Math}\ ,\ \bibinfo {pages}
  {82}} (\bibinfo {year} {1998})}\BibitemShut {NoStop}%
\bibitem [{\citenamefont {Kato}(1995)}]{Kato:1995}%
  \BibitemOpen
  \bibfield  {author} {\bibinfo {author} {\bibfnamefont {T.}~\bibnamefont
  {Kato}},\ }\href@noop {} {\emph {\bibinfo {title} {{Perturbation Theory for
  Linear Operators}}}}\ (\bibinfo  {publisher} {Springer, Berlin, Heidelberg},\
  \bibinfo {year} {1995})\BibitemShut {NoStop}%
\bibitem [{\citenamefont {Gitman}\ \emph {et~al.}(2012)\citenamefont {Gitman},
  \citenamefont {Tyutin},\ and\ \citenamefont {Voronov}}]{gitman:2012}%
  \BibitemOpen
  \bibfield  {author} {\bibinfo {author} {\bibfnamefont {D.}~\bibnamefont
  {Gitman}}, \bibinfo {author} {\bibfnamefont {I.}~\bibnamefont {Tyutin}},\
  and\ \bibinfo {author} {\bibfnamefont {B.}~\bibnamefont {Voronov}},\
  }\href@noop {} {\emph {\bibinfo {title} {Self-adjoint Extensions in Quantum
  Mechanics: General Theory and Applications to Schrödinger and Dirac
  Equations with Singular Potentials}}}\ (\bibinfo  {publisher} {Birkhäuser},\
  \bibinfo {year} {2012})\BibitemShut {NoStop}%
\bibitem [{\citenamefont {Kowalczyk}\ and\ \citenamefont
  {Paw\l{}owski}(2023)}]{Kowalczyk:2022ajp}%
  \BibitemOpen
  \bibfield  {author} {\bibinfo {author} {\bibfnamefont {M.}~\bibnamefont
  {Kowalczyk}}\ and\ \bibinfo {author} {\bibfnamefont {T.}~\bibnamefont
  {Paw\l{}owski}},\ }\bibfield  {title} {\bibinfo {title} {{Regularizations and
  quantum dynamics in loop quantum cosmology}},\ }\href
  {https://doi.org/10.1103/PhysRevD.108.086010} {\bibfield  {journal} {\bibinfo
   {journal} {Phys. Rev. D}\ }\textbf {\bibinfo {volume} {108}},\ \bibinfo
  {pages} {086010} (\bibinfo {year} {2023})},\ \Eprint
  {https://arxiv.org/abs/2212.12527} {arXiv:2212.12527 [gr-qc]} \BibitemShut
  {NoStop}%
\bibitem [{\citenamefont {Ashtekar}\ \emph {et~al.}(2010)\citenamefont
  {Ashtekar}, \citenamefont {Campiglia},\ and\ \citenamefont
  {Henderson}}]{Ashtekar:2010gz}%
  \BibitemOpen
  \bibfield  {author} {\bibinfo {author} {\bibfnamefont {A.}~\bibnamefont
  {Ashtekar}}, \bibinfo {author} {\bibfnamefont {M.}~\bibnamefont
  {Campiglia}},\ and\ \bibinfo {author} {\bibfnamefont {A.}~\bibnamefont
  {Henderson}},\ }\bibfield  {title} {\bibinfo {title} {{Path Integrals and the
  WKB approximation in Loop Quantum Cosmology}},\ }\href
  {https://doi.org/10.1103/PhysRevD.82.124043} {\bibfield  {journal} {\bibinfo
  {journal} {Phys. Rev. D}\ }\textbf {\bibinfo {volume} {82}},\ \bibinfo
  {pages} {124043} (\bibinfo {year} {2010})},\ \Eprint
  {https://arxiv.org/abs/1011.1024} {arXiv:1011.1024 [gr-qc]} \BibitemShut
  {NoStop}%
\bibitem [{\citenamefont {Flores-Gonz\'alez}\ \emph {et~al.}(2013)\citenamefont
  {Flores-Gonz\'alez}, \citenamefont {Morales-T\'ecotl},\ and\ \citenamefont
  {Reyes}}]{Flores-Gonzalez:2013zuk}%
  \BibitemOpen
  \bibfield  {author} {\bibinfo {author} {\bibfnamefont {E.}~\bibnamefont
  {Flores-Gonz\'alez}}, \bibinfo {author} {\bibfnamefont {H.~A.}\ \bibnamefont
  {Morales-T\'ecotl}},\ and\ \bibinfo {author} {\bibfnamefont {J.~D.}\
  \bibnamefont {Reyes}},\ }\bibfield  {title} {\bibinfo {title} {{Propagators
  in Polymer Quantum Mechanics}},\ }\href
  {https://doi.org/10.1016/j.aop.2013.05.005} {\bibfield  {journal} {\bibinfo
  {journal} {Annals Phys.}\ }\textbf {\bibinfo {volume} {336}},\ \bibinfo
  {pages} {394} (\bibinfo {year} {2013})},\ \Eprint
  {https://arxiv.org/abs/1302.1906} {arXiv:1302.1906 [math-ph]} \BibitemShut
  {NoStop}%
\bibitem [{\citenamefont {Qin}\ \emph {et~al.}(2012)\citenamefont {Qin},
  \citenamefont {Deng},\ and\ \citenamefont {Ma}}]{Qin:2012gaa}%
  \BibitemOpen
  \bibfield  {author} {\bibinfo {author} {\bibfnamefont {L.}~\bibnamefont
  {Qin}}, \bibinfo {author} {\bibfnamefont {G.}~\bibnamefont {Deng}},\ and\
  \bibinfo {author} {\bibfnamefont {Y.-G.}\ \bibnamefont {Ma}},\ }\bibfield
  {title} {\bibinfo {title} {{Path integrals and alternative effective dynamics
  in loop quantum cosmology}},\ }\href
  {https://doi.org/10.1088/0253-6102/57/2/28} {\bibfield  {journal} {\bibinfo
  {journal} {Commun. Theor. Phys.}\ }\textbf {\bibinfo {volume} {57}},\
  \bibinfo {pages} {326} (\bibinfo {year} {2012})},\ \Eprint
  {https://arxiv.org/abs/1206.1131} {arXiv:1206.1131 [gr-qc]} \BibitemShut
  {NoStop}%
\bibitem [{\citenamefont {Huang}\ \emph {et~al.}(2013)\citenamefont {Huang},
  \citenamefont {Ma},\ and\ \citenamefont {Qin}}]{Huang:2011es}%
  \BibitemOpen
  \bibfield  {author} {\bibinfo {author} {\bibfnamefont {H.}~\bibnamefont
  {Huang}}, \bibinfo {author} {\bibfnamefont {Y.}~\bibnamefont {Ma}},\ and\
  \bibinfo {author} {\bibfnamefont {L.}~\bibnamefont {Qin}},\ }\bibfield
  {title} {\bibinfo {title} {{Path Integral and Effective Hamiltonian in Loop
  Quantum Cosmology}},\ }\href {https://doi.org/10.1007/s10714-013-1520-2}
  {\bibfield  {journal} {\bibinfo  {journal} {Gen. Rel. Grav.}\ }\textbf
  {\bibinfo {volume} {45}},\ \bibinfo {pages} {1191} (\bibinfo {year}
  {2013})},\ \Eprint {https://arxiv.org/abs/1102.4755} {arXiv:1102.4755
  [gr-qc]} \BibitemShut {NoStop}%
\end{thebibliography}%

\end{document}